\documentclass[12pt]{article}
 \usepackage[nottoc]{tocbibind}
\usepackage[colorlinks=true,linkcolor=blue,urlcolor=magenta, citecolor=blue]{hyperref}
\usepackage{fullpage}
\usepackage{amssymb}
\usepackage{amsmath}
\usepackage[margin=0.5cm]{caption}
\usepackage{hyperref} 
\usepackage{cite}
\usepackage{upgreek}

\usepackage[pdftex]{graphicx}

\usepackage[utf8]{inputenc}
\usepackage[bb=boondox]{mathalfa}

\def\d{\partial}
\def\l{\left(}
\def\r{\right)}

\def\sign{\mbox{sign\,}}

\newcommand{\be}{\begin{equation}}
\newcommand{\ee}{\end{equation}}
\newcommand{\bea}{\begin{eqnarray}}
\newcommand{\eea}{\end{eqnarray}}
\newcommand{\bg}{\begin{gather}}
\newcommand{\eg}{\end{gather}}
\newcommand{\bseq}{\begin{subequations}}
\newcommand{\eseq}{\end{subequations}}

\def\Xint#1{\mathchoice
   {\XXint\displaystyle\textstyle{#1}}%
   {\XXint\textstyle\scriptstyle{#1}}%
   {\XXint\scriptstyle\scriptscriptstyle{#1}}%
   {\XXint\scriptscriptstyle\scriptscriptstyle{#1}}%
   \!\int}
\def\XXint#1#2#3{{\setbox0=\hbox{$#1{#2#3}{\int}$}
     \vcenter{\hbox{$#2#3$}}\kern-.5\wd0}}

\def\dashint{\Xint-}

\begin{document}
\baselineskip=15.5pt
\begin{titlepage}
\begin{center}
{\Large\bf Quantization of the Zigzag Model }\\
\vspace{0.5cm}
{ \large
John C. Donahue$^{a,b}$, Sergei Dubovsky$^a$
}\\
\vspace{.45cm}
{\small  \textit{   $^a$Center for Cosmology and Particle Physics,\\ Department of Physics,
      New York University\\
      New York, NY, 10003, USA}}\\ 
      \vspace{.25cm}
      {\small  \textit{   $^b$Kavli Institute for Theoretical Physics,\\
University of California, Santa Barbara, CA 93106
     }}\\ 

      \end{center}
\begin{center}
\begin{abstract}
The  zigzag model is a relativistic integrable $N$-body system describing the leading high-energy semiclassical dynamics on the worldsheet of long confining strings in massive adjoint two-dimensional QCD. We discuss  quantization of this model.  We demonstrate that to achieve a consistent quantization of the model it is necessary to account for the non-trivial geometry of  phase space. The resulting Poincar\'e invariant integrable quantum theory is a close cousin of $T\bar{T}$  deformed models. 
\end{abstract}
\end{center}
\end{titlepage}

%
%
%

\section{Introduction}

Quantization of a given classical system often feels more art than science \cite{faddeev2007complete}. There exists a range of prescriptions, from the very straightforward to the  heuristic and all the way to the highly technical and rigorous. A priori, none of them is guaranteed to work. Furtermore, if a quantization exists,  it does not have to be unique.
There is a good reason for all this; nature is intrinsically quantum. In general a more appropriate question to ask is what classical system(s) arises in the semiclassical regime(s) of a given quantum system, if such a regime can indeed be defined.
However, a human's quantum intuition is very limited. As a result, quantization of classical models is still one of the most efficient methods for constructing  interesting novel quantum systems.

 It helps that  the physics of the problem often imposes very restrictive  requirements on the admissible quantization.
This is exactly the situation considered in the present paper.  We show here how to quantize the zigzag model, which is a  relativistic $N$-body maximally superintegrable mechanical system recently identified in \cite{Donahue:2019adv,Donahue:2019fgn}. 
As we will see, even though the classical zigzag model is embarrassingly simple,  
it is surprisingly subtle to construct its proper quantization. 
The zigzag model describes $N$ massless particles on a line whose motion is governed by the following Hamiltonian
\be
\label{ZigzagH}
H=\sum_{i=1}^N|p_i|+\ell_s^{-2}\sum_{i=1}^{N-1}\l q_i-q_{i+1}+|q_i-q_{i+1}|\r\;.
\ee
The zigzag Hamiltonian was originally derived 
by considering the high-energy dynamics on the worldsheet of a confining string in two-dimensional adjoint QCD (with a single adjoint Majorana flavor) in the 't Hooft planar limit,
and $\ell_s^{-2}$ determines the tension of the confining flux tube in the fundamental representation. In what follows we set 
\[
\ell_s=1
\] 
unless specified otherwise.

The expectation is that it should be possible to set up a high energy expansion on the worldsheet using the zigzag  model as a leading order approximation. For this idea to be successful, the zigzag model itself needs to be Poincar\'e invariant and solvable. This is indeed the case at the classical level. The classical zigzag model is Poincare invariant, Liouville integrable ({\it i.e.}, it exhibits $N$ globally defined charges in involution) and maximally superintegrable ({\it i.e.}, it is possible to construct  $2N-2$ algebraically independent conserved charges in addition to the Hamiltonian).
Hence, in this paper we are looking for the quantization of the zigzag model which preserves both Poincar\'e invariance and integrability. 

Classical scattering in the zigzag model
gives rise to a time delay proportional to the collision energy. In the quantum language this time delay corresponds to the celebrated shock wave phase shift  \cite{'tHooft:1987rb,Amati:1987wq},
\be
\label{shock}
e^{i\delta}=e^{{is\over 4}}\;,
\ee
 which also describes the worldsheet scattering of critical strings \cite{Dubovsky:2012wk} and, more generally, scattering arising as a result of the $T\bar{T}$ deformation \cite{Dubovsky:2013ira,Smirnov:2016lqw,Cavaglia:2016oda}. This suggests that the classical zigzag model describes an $N$-particle subsector of the massless $T\bar{T}$-deformed fermion similarly to how the 
 Ruijsenaars--Schneider model \cite{Ruijsenaars:1986vq} describes an $N$-soliton subsector of the sine-Gordon model. Given that the $T\bar{T}$ deformation can be described in terms of the one-loop exact path integral \cite{Dubovsky:2018bmo} this connection suggests that an integrable quantization of the zigzag model results in the classical shock wave phase shift as an exact quantum answer.
 
 The relation to the $T\bar{T}$-deformed theories raises a number of interesting conceptual questions about the zigzag model. Indeed, the $T\bar{T}$-deformation describes a relativistic quantum filed theory 
 coupled to a topological gravity \cite{Dubovsky:2017cnj,Cardy:2018sdv,Dubovsky:2018bmo}. As a result one does not expect the existence of local off-shell observables in $T\bar{T}$ deformed models. At first sight this seems to be at odds with the relation between the $T\bar{T}$ deformation and the zigzag model. Indeed, one may expect that the positions of particles in the quantum zigzag model  provide a set of well-defined local off-shell observables. 
 However, this expectation is somewhat too naive.  
  Indeed, this subtlety is well familiar already from a first quantized description of conventional free relativistic particles.
  In particular, amplitudes defined by the relativistic invariant path integral
  \be
  \langle x|y\rangle=\int{\cal D} Xe^{iS_{pp}[x,y]}\;,
  \ee
   cannot be interpreted as conventional transition amplitudes in the position space. Instead, they correspond to the Feynmann propagator of the second quantized field theory.
    We will see that the zigzag case is even more subtle.
  Our quantization of the zigzag model indeed reproduces the $T\bar{T}$ $S$-matrix at the quantum level.
  However, at present it is unclear whether it leads to well-defined off-shell observables.

The structure of the paper is as follows. In section \ref{sec:naive} we present the naive straightforward quantization of the zigzag model in the two-particle case $N=2$. We will see that this approach does
not lead to a satisfactory quantization of the zigzag model. In section \ref{sec:Zigzag} we trace the problem to the non-trivial phase space geometry of the model, which is ignored by the naive quantization. We argue that a consistent quantization of the zigzag model is achieved by making use of the globally defined action angle variables constructed in \cite{Donahue:2019fgn}.
In section \ref{sec:Smatrix} we explain how to reconstruct the $T{\bar T}$ $S$-matrix (\ref{shock}) using this quantization.
The construction is very similar to the $T{\bar T}$ $S$-matrix derivation as presented in \cite{Dubovsky:2017cnj}. Namely,  action angle variables provide a formulation of the zigzag model in terms of free particles. The non-trivial $S$-matrix (\ref{shock}) arises as a consequence of introducing ``dynamical" physical coordinates, which are different from the free ones. In section \ref{sec:mult} we briefly describe the extension of all these results to the multi-particle case. In section \ref{sec:thooft} we comment on the relation to the closed ``folded" strings \cite{Bardeen:1975gx,Bardeen:1976yt,Hanson:1976ey} and the 't Hooft model \cite{tHooft:1974pnl}. 
We conclude in section~\ref{sec:last}.

%

\section{Failures of the Naive Quantization}
\label{sec:naive}

The zigzag Hamiltonian in the two-particle subsector is given by 
\be
\label{eq:2ham}
H = |p_1| + |p_2| + q_1 - q_2 + |q_1 - q_2| \,\,\,.
\ee
At first sight it is natural to quantize this model using the standard canonical quantization prescription
\be
\label{canonical}
[q_i,p_j]=i\delta_{ij}\;.
\ee
In the coordinate presentation one then encounters a somewhat unconventional operator $|p|$.
However, it   is  straightforward to define it
via the Fourier transform,
\[
|p|=-\int{dq\over 2\pi}e^{ikq}\l{1\over(q-i\epsilon)^2}+{1\over(q+i\epsilon)^2}\r\equiv -\int{dq\over \pi}e^{ikq}{{\cal P}\over q^2}
\]
so that 
\[
|p|\psi(q)=-{1\over \pi} \int_{-\infty}^{\infty} dq'\psi(q',t)\frac{{\cal P} }{(q-q')^2} \;.
\] 

The first indication that this quantization is problematic stems from the following observation. An important step in the classical analysis of the integrable structure of the zigzag model presented in 
\cite{Donahue:2019adv,Donahue:2019fgn} is the construction of a conserved topological charge $T$,
which in the two-particle case reduces to 
\be
\label{T}
T={1\over 2}\l-s_1+s_1s_{1,2}+s_{1,2}s_2-s_2 \r\;,
\ee
where
\[
s_i=\sign p_i\;,\;\;s_{i,j}=\sign (q_i-q_j)\;.
\]
The existence of this topological invariant implies that the classical phase space splits into three distinct superselection sectors, labeled by the possible values of $T=-2,0,2$. These values simply count the number of  left- and right-movers in the asymptotic regions $t\to \pm \infty$\footnote{At intermediate times they count the number of left- and right-movers ``along the string worldsheet" \cite{Donahue:2019fgn}.}.

Of course, for two particles the existence of these sectors is completely obvious.
The values $T=\pm2$  correspond to the $LL$  ($RR$) sectors describing  two left(right)-moving particles which stay free at all times. The value $T=0$ gives rise to the only interacting $LR$ sector present in the two-particle case.

It is immediate to see that the free $LL$ and $RR$ sectors are lost with the naive quantization. Indeed, states in the $LL$ sector can be characterized by the condition
\be
\label{LLcond}
(P+H)\psi(q)=0\;,
\ee
where 
\[
P=p_1+p_2
\]
is the total momentum.
This condition implies that the wave function has to satisfy 
\begin{gather}\label{p1cond}
(p_1+|p_1|)\psi(q)=0\\
(p_2+|p_2|)\psi(q)=0\\
\label{qcond}
(q_1-q_2+|q_1-q_2|)\psi(q)=0\;.
\end{gather}
The first condition (\ref{p1cond}) implies that 
\[
\psi(q_1,q_2)=\int_0^\infty dp_1 e^{ip_1q_1}f(p_1,q_2)
\]
is an analytic function of $q_1$ in the upper half plane $\mbox{\rm Im }q_1>0$. Hence, it cannot vanish at all values $q_1<q_2$ as required by the last condition (\ref{qcond}).

One might try to get around this difficulty by postulating that one should first restrict to a certain classical subsector before performing the quantization, so that the naive quantization describes the $LR$ sector only. The free  $RR$ and $LL$ sectors are straightforward to construct separately---these describe a pair of free massless particles with positive and negative momenta.

\begin{figure}
    \centering
    \textbf{}\par\medskip
    \includegraphics[scale=0.4]{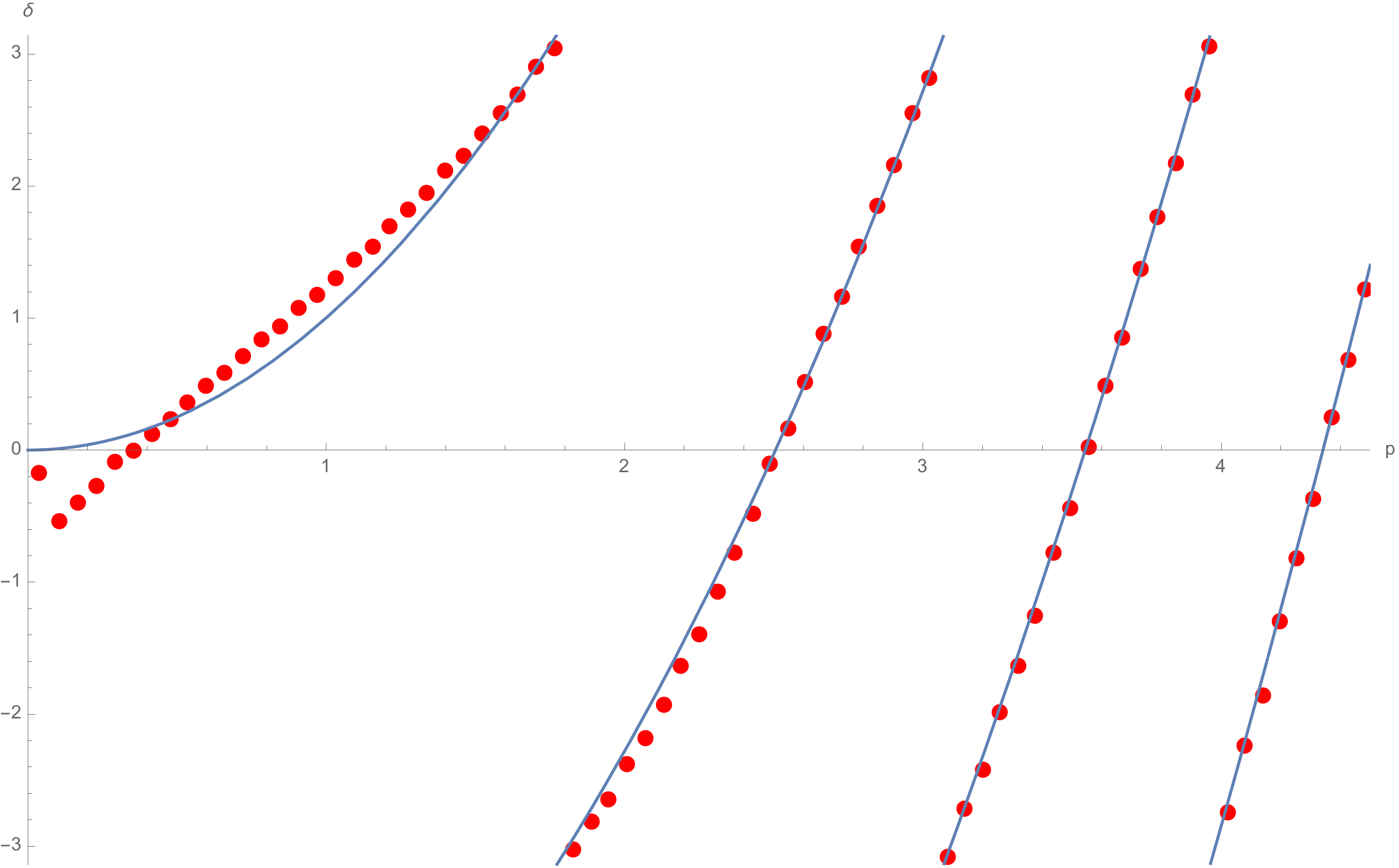}
    \caption{Numerical phase shift corresponding to the Schr\"odinger equation (\ref{eq:SE}) (red points) versus the shock wave phase shift (solid lines).}
    \label{fig:phase}
\end{figure}

To see that this still does not lead to a satisfactory quantization of the zigzag model, let us inspect the resulting Schr\"odinger equation in the $P=0$ frame,
\be
\label{eq:SE}
   i \d_{t} \psi(q,t)  = - \frac{2}{\pi} \int_{-\infty}^{\infty} dq'\psi(q',t)\frac{{\cal P} }{(q-q')^2} + (q + |q|)\psi(q,t) \,,
\ee
and calculate the corresponding scattering phase shift. Note that as a consequence of the non-local nature of the $|p|$ kernel, one cannot proceed by solving this equation in $q>0$ and $q<0$ regions and gluing the solutions at $q=0$, as one would had done in the conventional quantum mechanics (see, e.g. \cite{jeng2010nonlocality}).
As far as we can tell, this equation cannot be solved analytically, so we resort to the numerical determination of the scattering phase shift. We discuss details of this calculation in the
Appendix~\ref{app:numerics}. The result is presented in Fig.~\ref{fig:phase}.
One observes that in the semiclassical (large momentum) regime the naive quantization perfectly agrees with the classical shock wave phase shift (\ref{shock}). However, the two phase shifts disagree at intermediate momenta $p\sim 1$, even though they remain quite close to each other at almost all momenta\footnote{Numerical results cannot be trusted at $p\ll 1$ due to numerical finite size effects. However, we checked that the disagreement at $p\sim 1$ is not caused by the numerics.}.
This demonstrates that the classical equivalence between the zigzag model and the $T\bar{T}$ deformation  does not extend at the quantum level if one
follows the  naive quantization. 

Perhaps the most severe trouble with the naive quantization becomes manifest upon the inspection of the Poincar\'e algebra in the zigzag model. At the classical level the boost symmetry generator takes the following form,
\be
\label{boost}
J=\sum_{i=1}^Nq_i|p_i|+{1\over 2}\sum_{i=1}^{N-1}(q_i+q_{i+1})(q_i-q_{i+1}+|q_i-q_{i+1}|)\;.
\ee
At the level of classical Poisson brackets this generator forms the $ISO(1,1)$ Poincar\'e algebra together with the Hamiltonian $H$ and the total momentum $P$. However, at the quantum level the algebra is spoiled by  contact terms, which appear uncurable (this observation in a related  model has also been made in \cite{Artru:1983gm,Lenz:1995tj}).


\section{Geometry of the Phase Space and Quantization }
\label{sec:Zigzag}
Failures of the naive quantization described in the previous section appear as a set of disconnected technical issues. To construct a successful alternative quantization it is important to find an underlying general reason for these shortcomings. We argue here that they all are related to the non-trivial phase space geometry of the zigzag model which is completely ignored by the naive 
quantization.

To see the origin of this non-trivial geometry it is convenient to separate the bulk and the relative motion by performing the following canonical coordinate change,
\begin{gather}
P=p_1+p_2\;,\;\; \bar{q}={q_1+q_2\over 2}\\
p={p_1-p_2\over 2}\;,\;\; q=q_1-q_2\;,
\end{gather}
so that the Hamiltonian (\ref{eq:2ham}) turns into
\be
\label{H2new}
H=\left|{P\over 2}+p\right|+\left|{P\over 2}-p\right|+q+|q|\;.
\ee

\begin{figure}
    \centering
    \textbf{}\par\medskip
    \includegraphics[scale=0.4]{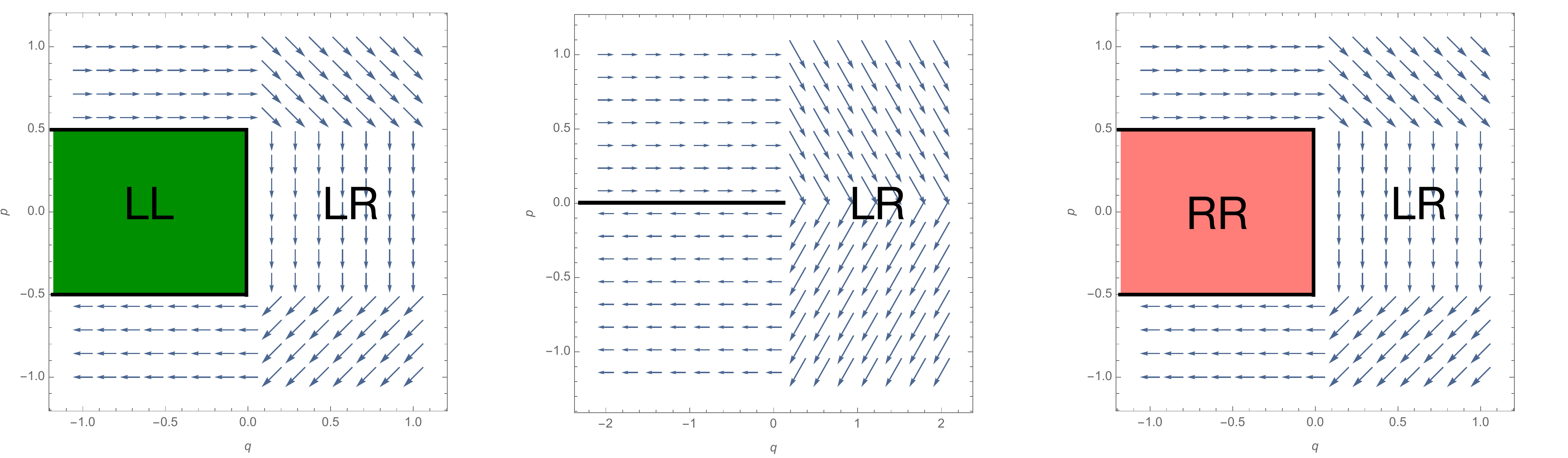}
    \caption{Hamiltonian flows of the zigzag model in the $(p,q)$ plane at $P=-1$ (left panel), $P=0$ (middle panel) and $P=1$ (right panel). 
    Topological sectors corresponding to different values of the charge $T$ are labeled as $LL$, $RR$ and $LR$.
    The vector field vanishes in the free $LL$ and $RR$ regions.}
    \label{fig:phasespace}
\end{figure}
In Fig.~\ref{fig:phasespace} we presented phase portraits of the zigzag model in the $(q,p)$ plane at positive, zero and negative values of the total momentum $P$. One immediately finds that the Hamiltonian vector flow of the zigzag model is badly discontinuous at the boundaries of the topological sectors corresponding to different values of the topological charge (\ref{T}). Note that the vector field itself has additional discontinuities inside the $LR$ region. However, these can easily be smoothed out and the geometry of the flow lines is continuous there, unlike at the boundaries between the topological sectors. This strongly suggests that the proper classical phase space of the $N=2$ zigzag model is not the
full $\mathbb{R}^4$, but one needs to exclude these boundaries. Equivalently, one needs to quantize in each of the sector separately, 
accounting for the fact that the sectors are non-trivial subregions in $\mathbb{R}^4$, which is ignored by the naive quantization of Section~\ref{sec:naive}.

This point gets even stronger when the Poincar\'e invariance of the zigzag model is taken into account. Indeed, the flow generated by the boost generator changes the value of the total momentum $P$. 
However, the sign of $P$ does not change in the free $LL$ and $RR$ sectors, so that $P$ stays positive in $RR$ and negative in $LL$. On the other hand, by applying the boost in the $LR$ sector one may change the value of $P$ (including its sign) arbitrarily. Consequently, also the classical flow corresponding to the boost generator $J$ is badly discontinuous at the boundaries between different sectors.
 This explains why the naive quantization, which ignores the phase space geometry, is incompatible with the Poincar\'e symmetry.

\subsection{Free $LL$ and $RR$ sectors} 
Let us now describe a consistent quantization of the zigzag model guided by these geometrical considerations.
The most straightforward way to exclude the boundaries between the topological sectors from $\mathbb{R}^4$ is to perform quantization in each of the sectors separately.
Let us start with the free $LL$ and $RR$ sectors. At first sight these are completely trivial, however, even here we encounter a subtlety.
For concreteness, let us focus on the $RR$ sector. Here the phase space is a subregion of $\mathbb{R}^4$ determined by the following inequalities,
\begin{gather}
p_1>0\;,\;\; p_2>0\;,\\
q_1-q_2>0\;,
\label{qpos}
\end{gather}
which is the same as $\mathbb{R}_+^3\times \mathbb{R}$. The Hamiltonian is simply
\[
H_{RR}=p_1+p_2\;.
\]
If the range of coordinates $q_1$, $q_2$ were not restricted, {\it i.e.}, if the phase space geometry were $(\mathbb{R}_+\times \mathbb{R})^2$, the quantization would be straightforward.
The corresponding Hilbert space is spanned by the momentum eigenstates $|p_1,p_2\rangle$ with positive momenta $p_1,p_2>0$. Note, however, that a particle on a half-line (or, equivalently, a particle with a positive momentum) provides perhaps the simplest example where a non-trivial phase space geometry ($\mathbb{R}_+\times \mathbb{R}$ in this case) has important consequences. This example is often used as a testing ground for more sophisticated quantization methods, such as group theoretical quantization \cite{Isham:1983zr}.

The subtlety is that the coordinate operator cannot be extended to a self-adjoint operator if the range of  momenta is restricted to a half-line.
In particular, no coordinate represenation exists for the Hilbert space. Hence it is  problematic to enforce the inequality (\ref{qpos}) in the $RR$ sector. This complication is a direct consequence of the uncertainty principle, and essentially equivalent to the reason why the naive quantization misses the $LL$ and $RR$ sectors, as articulated in the beginning of section \ref{sec:naive}.

We believe that the most natural way of getting around this problem is to declare that particles in the zigzag model are identical in the $RR$ and $LL$ sectors. At the classical level  this
amounts to identifying points in the phase space related by particle permutation
\[
(p_1,q_1,p_2,q_2)\sim(p_2,q_2,p_1,q_1)\;.
\]
 This automatically enforces the  constraint (\ref{qpos}) (or, better to say, restricts $(q_1-q_2)$ to be on a half-line). At the quantum level this is implemented by imposing the identification
\[
|p_1,p_2\rangle=\pm |p_2,p_1\rangle\;,
\]
where the sign determines  bosonic or fermionic statistics, as usual.
This prescription amounts to a non-trivial modification of the classical zigzag model.  Note that the statistical identification cannot be imposed in the $LR$ sector, because the zigzag Hamiltonian is not invariant under the particle exchange there. 

It is somewhat unconventional that the statistics in the zigzag model is imposed only in certain subsectors of a theory. 
However, it is in fact quite natural 
 when the zigzag model is obtained as a high energy limit of the adjoint QCD.
This is related to the off-shell particle identity arising as a consequence of color ordering in the worldsheet theory, as discussed in \cite{Dubovsky:2018dlk}. As explained there, 
worldsheet excitations correspond to identical particles only at the level of asymptotic scattering states. The free $LL$ and $RR$ sectors are in a sense always on-shell, so that the statistical identification can (and should) be imposed directly there. On the other hand, it is impossible to impose particle identification off-shell  in the $LR$ sector, where zigzags forms during scattering processes. Conventional quantum statistics gets restored in this sector at the level of scattering states, because the potential in (\ref{H2new}) has only one flat asymptotic region $q<0$, as if scattering were taking place on a half-line.

\subsection{Interacting $LR$ sector}
\label{sec:interactingLR}
Let us now come to the quantization of the dynamical $LR$ sector. The discussion above strongly suggests that a successful quantization can be achieved by switching to a set of phase space coordinates that are better adjusted to the geometry of the $LR$ region.
Given that we would like to preserve classical integrability at the quantum level, it is natural to follow the classical integrable structure as a guide for the quantization. 

As demonstrated in \cite{Donahue:2019fgn}, the zigzag model with $N$ particles exhibits
 $2N-1$ independent integrals of motions. In the asymptotic $t\to\pm\infty$ regions these reduce to the  particle momenta and pairwise interparticle separations between  particles moving in the same direction. For $N=2$ the integrals are 
\begin{gather}
\label{IR}
   P_1={P+H\over 2}\\
    \label{IL}
   P_2={P-H\over 2}\\
    \label{Pt}
    \tilde{P} =p_2 s_2-\frac{q_1}{2}\left(1-s_{1,2}\right)-\frac{q_2}{2} \left(3+ s_{1,2}\right) \,\,\, .
\end{gather}
These expressions are simpler than those provided in the Appendix A of  \cite{Donahue:2019fgn}. The reason is that here we simplified expressions for the charges, using that we restrict to the $LR$ sector only. Note that the first two integrals (\ref{IR}), (\ref{IL}) are translationally invariant, while the last one may be thought of as a dynamical ``rod" variable---it shifts linearly under an overall shift of the particle positions.
In addition to these integrals,  a natural ``clock" variable --- a quantity which depends linearly on time when equations of motion are satisfied---was constructed in \cite{Donahue:2019fgn},
\be
\label{Ht}
\tilde{H}=p_2+{q_1-q_2\over 2}s_1(1-s_{1,2})\;.
\ee

It is straightforward to see that $(P_1,P_2,\tilde{P},\tilde{H})$ define a globally well-defined parametrization of the $LR$ region of the phase space (momentarily, we will provide an explicit inverse mapping to the $(p,q)$ variables). Hence it is natural to use these coordinates as a basis for quantization. To be precise, let us define the following coordinate variables 
\begin{gather}
\label{eq:freepos}
Q_1={1\over 2}\l \tilde{H}-\tilde{P}-P_2\r\,,
\\
Q_2={1\over 2}\l-\tilde{H}-\tilde{P}+P_1\r\,,
\end{gather} 
%
%
%
%
which, together with momenta $(P_1,P_2)$ form a global set of canonical coordinates in the $LR$ sector.

In these coordinates, the zigzag Hamiltonian (\ref{eq:2ham}) takes a simple form
\be
 H = P_1 - P_2 \, .
\ee
In fact, the whole Poincare algebra  rewritten in terms of these coordinates takes the free particle form 
\begin{gather}
\label{Ptot}
 P = P_1 + P_2 \\
 \label{Boost}
 J = Q_1 P_1 - Q_2 P_2
\end{gather}
and
\be
\{ H,P\}=0\;,\;\;\{J,P\}= H\;,\;\;\{J, H\}=P\;.
\ee
Hence, $(P_1,P_2,Q_1,Q_2)$ is a set of action-angle variables for the zigzag model. Importantly,  these action-angle variables  are globally well-defined---$(P_1,P_2,Q_1,Q_2)$ provide a one-to-one parametrization of the $LR$ sector, provided the momenta are restricted to a half-line
\be
P_1>0\;,\;\;P_2<0\;,
\ee
as follows from (\ref{IR}), (\ref{IL}).
Indeed, by fixing the values of the conserved quantities $P_1$, $P_2$ and 
\[
\bar{Q}={Q_1+Q_2\over 2}
\]  
one uniquely determines the phase space trajectory, and then the remaining ``clock" variable 
\[
Q=Q_1-Q_2
\] picks a point on the trajectory. 

In addition to this indirect argument it is also straightforward to explicitly reconstruct the original physical coordinates through the action-angle variables. Indeed, $(Q,P)$ and $(q,p)$
variables are piecewise linearly related to each other, with the exact form of the relation being determined by the values of $s_1$, $s_2$ and $s_{1,2}$. So for each possible value of  $s_1$, $s_2$ and $s_{1,2}$\footnote{Note that the values $s_1=s_2=s_{1,2}=\pm 1$ are not possible because these correspond to $LL$ and $RR$ sectors. } one may solve for $(q,p)$'s in terms of $(Q,P)$'s. After this is done, one can  rephrase the choice of $s_1$, $s_2$, $s_{1,2}$ in terms of $(Q,P)$ variables. 

The result of this procedure can be summarized by the following expressions,
\begin{gather}
\label{q}
q={H\over 2}-{|Q|\over 2}-{|P|\over 4}-{1\over 2}\left||Q|-{|P|\over 2}\right|\\
\label{p}
p={\mbox\sign(Q)\over 2}\l\left|{H\over 2}-|Q| \right|-{H\over 2}-|Q|\r\\
\label{qbar}
\bar{q}=\bar{Q}+{\sign(Q)\sign(P)\over 4}\l {|P|\over 2}+|Q|-\left|{|P|\over2}-|Q|\right|\r\;.
\end{gather}

%
%
%
%
%
%
In the new canonical variables the $N=2$ zigzag model turns into a system of 
 two massless  free particles with each of momenta restricted to a half-line, so that the phase space geometry is  $ (\mathbb{R} \times \mathbb{R}^+) \times (\mathbb{R} \times \mathbb{R}^-)$. Quantization of this phase space is most straightforward  to 
 perform in the momentum representation, so that the Hilbert space is spanned by the vectors 
 \[
 |P_1,P_2\rangle\;.
 \]
 As we already mentioned, coordinate operators
 \[
 Q_i=i\d_{P_i}
 \]
do not admit self-adjoint extension for this system. Geometrically, this can be traced to the fact that the Hamiltonian flow corresponding to the coordinate operator does not map the half-line of positive momenta into itself\footnote{Note that for certain observables with this property, such as $\hat{Q}_i^2$,  one can still define a self-adjoint operator by  introducing appropriate boundary conditions at $P_i=0$. More generally, this can always be done for positive definite operators using the Friedrichs extension.
}. Still, for all practical purposes a theory of a particle with positive momentum is perfectly local. Indeed, given a function $f(P)$, such that 
 \[
 f(0)=0\;, \;\;f(\infty)=1
 \]
 one may construct a regularized essentially self-adjoint coordinate operator
 \be
 \label{regQ}
\hat{ Q}_{i,f}=f(\hat{P}_i) \hat{Q}_i f(\hat{P}_i)\;.
 \ee
 By considering a family of functions $f$, which approach unity almost at all values of momenta apart from a small vicinity of the origin $P=0$, one obtains a family of regularized operators $\hat{Q}_{i,f}$ whose action on wave packets carrying non-zero momenta approximates the coordinate operators $\hat{Q}_i$ with any desired precision. 
Note that the Lorentz boost generator (\ref{Boost}) does not require any regularization and that boosts act as
 \[
  |P_1,P_2\rangle\to |\lambda P_1,\lambda^{-1}P_2\rangle\;.
 \]
 
 At first sight the existence of these globally defined action-angle variables turns the zigzag model into a free system, however this is not the case. 
 To properly describe the physics one needs to get back to the original coordinates as defined by (\ref{q}), (\ref{p}), (\ref{qbar}). Note that this situation is not at all unusual. Recall, that according to Darboux's theorem (for contact forms) any mechanical system can be brought into a canonical free form by a change of coordinates  in a vicinity of a generic point. Consequently, at least locally, all physical content of a given system is determined by a coordinate choice.
 
 At the quantum level it is not immediately obvious that the rather complicated looking expressions (\ref{q}), (\ref{p}), (\ref{qbar}) allow us to unambiguously define the corresponding quantum operators. 
 However,  we will see now that using regularized versions of the physical $(\hat{p},\hat{q})$ operators it is possible to construct at least some quantum observables, such as the $S$-matrix.
 
 \section{Two-particle $S$-matrix}
 \label{sec:Smatrix}
 \subsection{Scattering in the rest frame}
 \label{subsec:restframe}
 Let us now use this quantization to derive the exact $S$-matrix in the $LR$ sector. For simplicity,  let us first consider scattering in the rest frame $P=0$. Then our quantization lands us in the energy representation, with the $P=0$ basis states of the form
   \[
 |H/2,-H/2\rangle\equiv|H\rangle\;.
 \]
The expression for the relative coordinate
 \[
\hat{ q}=\hat{q}_1-\hat{q}_2\;,
 \] 
 as determined by (\ref{q}), simplifies in the rest frame to 
 \be
 \label{qrest}
 \hat{q}={\hat{H}\over 2}-\hat{|Q|}\;.
\ee
Here, as before,
\[
\hat{Q}=2i\d_H\;.
\]
This operator can be thought of as a ``clock" operator in the following sense. Let us consider a  $\hat{Q}$ eigenstate with eigenvalue $Q$,
\be
\label{Qhat}
|Q\rangle=\int dH e^{-iH Q/2}|H\rangle\;.
\ee
Its time evolution amounts to a shift $Q\to Q+2 t$, so that performing measurements of $\hat{Q}$ is equivalent to measuring time $t$.

To calculate the $S$-matrix we follow the standard prescription of stationary scattering theory. Namely, we deduce the phase shift from the behavior of a stationary wave function in the coordinate representation,
\be
\psi_H(q)=\langle q|H \rangle
\ee
in the free region $q\to-\infty$. Importantly, we are using the physical coordinate $q$ here.
By making use of (\ref{qrest}) we find that $\psi_H(q)$ satisfies the following equation,
\be
\label{regqeq}
q\psi_H(q)={H\over 2}\psi_H(q)-\int_0^\infty dH_1\psi_{H_1}(q)\langle H_1||\hat{Q}||H\rangle\;.
\ee
Given that the clock operator $\hat{Q}$ does not have a self-adjoint extension, in order to define an operator $|\hat{Q}|$  we use a regularized operator $\hat{Q}_f$, which takes  form 
\be
\label{Qf}
\hat{Q}_f=2i f^2(H)\d_H+2if(H)\d_H f(H)
\ee
in the $H$ representation. This operator has a continuum spectrum $Q\in(-\infty,\infty)$. Using (\ref{regQ}) one finds that the corresponding eigenfunctions take the following form,
\be
|Q\rangle_f=\int_0^\infty {dH \over 2\sqrt{\pi}f(H)} e^{-iQ\int^H{ d\tilde{H}\over 2f^2(\tilde{H})}}|H\rangle\;,
\ee 
which are normalized as
\[
_f\langle Q_1|Q_2\rangle_f=\delta(Q_1-Q_2)\;.
\]
This allows us to define a regularized  matrix element of $|\hat Q|$ as
\[
\langle H_1||\hat{Q}_f||H\rangle={1\over 4\pi}\int_{-\infty}^{\infty} {dQ \over  f(H)f(H_1)}|Q|e^{iQ\int_H^{H_1}{ d\tilde{H}\over 2f^2(\tilde{H})}}=-{2\over \pi f(H)f(H_1)}\mbox{\rm Re}
 \l \int_H^{H_1}{ d\tilde{H}\over f^2}+i\epsilon\r^{-2}\;.
\]
At this stage it is natural to remove the regularization by setting $f(H)=1$. As a result, (\ref{regqeq}) turns into the following equation for the physical stationary wave function,
\be
\label{qeq}
q\psi_H(q)={\ell_s^2\over 2}H\psi_H(q)+{2\over \pi}\int_0^\infty dH_1\psi_{H_1}(q){{\cal P}\over (H_1-H)^2}\;,
\ee
where we restored the explicit dependence on the string tension $\ell_s$.
We see that the sole role of the regularization is to motivate the definition of the $|Q|$ operator.

 Note that the eigenvalue equation (\ref{qeq}) is quite different from a stationary Schr\"odinger equation that arises in conventional scattering theory. Namely, it is written in the energy representation, so that the roles played by the coordinate $q$ and energy $H$ are interchanged---the coordinate $q$ enters in (\ref{qeq}) as an eigenvalue. Related to this, (\ref{qeq}) is linear w.r.t. multiplying by an arbitrary function of $q$, but not by a function of $H$. Operationally this happened because we already fixed relative phases of energy eigenstates $|H\rangle$  by defining $\hat{Q}$ via (\ref{Qhat}).
 
 To determine the scattering phase shift we need to solve (\ref{qeq}) in the free region $q\to -\infty$. One expects the wave function
 to turn into a sum of an incoming and scattered waves there,
\be
\psi_H(q)|_{q\to-\infty}=\psi_++\psi_-\;,
\ee
where
\be
\label{expected}
\psi_{\pm}\sim A_\pm e^{\pm {i\over 2} q H+i\delta_\pm(H)}\;.
\ee
To determine the phase shifts $\delta_\pm(H)$ let us make use of the derivative of the Sokhotski formula
\be
\label{Sokhot}
{{\cal P}\over (H_1-H)^2}={1\over( H_1-H\mp i\epsilon)^2}\pm i\pi\delta'(H_1-H)\;.
\ee
Namely, note that at $q\to -\infty$ the scattered wave $\psi_+$ is exponentially small in the lower half-plane, $\mbox{\rm Im } H<0$ and the incoming wave $\psi_-$ is exponentially small in the upper half plane $\mbox{\rm Im } H>0$. Then by using the upper sign in (\ref{Sokhot}) for $\psi_+$ and the lower one for $\psi_-$ we may Wick rotate the integration contours by $\mp{\pi/2}$  without encountering any singularities. As a result,  (\ref{qeq}) takes the following form 
\be
\label{rotatedqeq}
 \ell_s^2{H\over 2}(\psi_++\psi_-)+2\delta_+'\psi_+-2\delta_-'\psi_-+A_+I_++A_-I_-=0
\ee
where
\be
\label{Ipm}
I_\pm(H)=\mp {2\over \pi}\int_0^\infty dh {1\over\l  ih\pm H+i\epsilon\r^2}e^{qh+i\delta_\pm(\mp ih)}\;.
\ee
At $H\neq 0$ these integrals vanish in the $q\to -\infty $ region, which allows us to determine the phase shifts from (\ref{rotatedqeq})
\be
\delta_\pm=\mp\ell_s^2 {H^2\over 8}+c_\pm\;.
\ee
where $c_\pm$ are $H$-independent integration constants. This approximation is valid provided
\be
\label{range}
H\gg{1\over (q\ell_s^2)^{1/3}}\gg{1\over q}\;.
\ee
Indeed, in this range 
 integrals in (\ref{Ipm}) can be estimated as $I_{\pm}\sim {1\over qH^2}$, which can be neglected compared to other terms in (\ref{rotatedqeq}).

It is worth noting that for these phase shifts $e^{i\delta_\pm}$ is exponentially small at large values of $|H|$ in the fourth quadrant of the complex plane, and $e^{i\delta_\pm}$ is exponentially small
in the first quadrant. Hence, to justify the Wick rotation which we performed, one does not need to take the strict $q\to-\infty$ limit---it can be performed at large finite negative $q$ as well.

Note that at this stage we still have a freedom to multiply $\psi_\pm$ by arbitrary functions of $q$.
In other words, up to now we determined that the wave function  in the $q\to-\infty$ region takes form
\be
\label{psiA}
\psi_H(q)|_{q\to-\infty}=A_+(q)e^{{i\over 2}qH-{i\over 8}H^2}+A_-(q)e^{-{i\over 2}qH+{i\over 8}H^2}\;.
\ee
To fix the $A_\pm$ amplitudes let us inspect the equation (\ref{qeq}) in the small $H$ limit. Here one may neglect the first term on the r.h.s. so that the equation reduces to 
\be
\label{qeq0}
q\psi_H(q)={2\over \pi}\int_0^\infty dH_1\psi_{H_1}(q){{\cal P}\over (H_1-H)^2}\;.
\ee
This equation corresponds to the infinite tension, $\ell_s=0$, limit of the zigzag model. Previously, this integral equation appeared in the semiclassical analysis of the 't Hooft equation \cite{Brower:1978wm,Fateev:2009jf} and can be solved exactly. We review this solution in  Appendix~\ref{app:Mellin}.
The resulting solution takes the following form in the $qH\to-\infty$ limit,
\be
\label{psi0}
\psi^{0}_H(q)|_{q\to-\infty}=e^{{i\over 2}qH+i{3\pi\over 8}}+e^{-{i\over 2}qH-i{3\pi\over 8}}\;.
\ee
This approximation is valid in the range
\[
{1\over q}\ll H\ll {1\over \ell_s}\;,
\]
 which overlaps with (\ref{range}). Then by requiring  that the two approximations (\ref{psiA}) and (\ref{psi0}) match in the overlap region one finds  that
\[
A_{\pm}(q)=e^{\pm i{3\pi\over 8}}\;.
\]
As a result, the scattering wave function in the free asymptotic region determines the  phase shift to be
\be
\label{phaseshift}
\delta\equiv \delta_--\delta_+={\ell_s^2H^2\over 4}-{3\pi\over 4}\;,
\ee
which reproduces the $T\bar{T}$ phase shift (\ref{shock}) up to a constant $-3\pi/4$ shift.
%

Note that the energy dependent part of the phase shift can be obtained by using  a quicker argument, which is parallel to the classical one presented in  \cite{Donahue:2019fgn}.
The argument  again relies on the relation (\ref{qrest}) and makes a direct use of $\hat{Q}$ as a clock variable. Namely, let us consider a wave packet peaked around $q(t)$ in the physical coordinate space and around $H$ in energy.
Then  from  (\ref{qrest}) one finds that at early, $t_e\to-\infty$, and late, $t_l\to\infty$, times  
\begin{gather}
\label{early}
q(t_e)={H\over 2}+Q(t_e)\\
\label{late}
q(t_l)={H\over 2}-Q(t_l)\;,
\end{gather}
where 
\[
Q(t)=Q_0+2t
\]
is the trajectory of the wave packet in the $Q$ space. By taking the sum of (\ref{early}) and (\ref{late}) one finds that
\be
t_l-t_e=-{q(t_e)+q(t_l)\over 2}+{H\over 2}\;,
\ee
which corresponds to the time delay $H/2$ in agreement with the phase shift (\ref{phaseshift}). One may be worried though that this argument is not rigorous enough given that strictly speaking
the $Q$ representation does not exist in this setup because $\hat{Q}$ is not a self-adjoint operator. A more rigorous and detailed derivation presented above gives  confidence that this issue is mostly a technicality, and provides a tractable description of the quantum scattering process directly in the physical coordinates via the scattering equation (\ref{qeq}). 

\subsection{Scattering in a general frame}
The quantization described in section~\ref{sec:Zigzag} is manifestly Poincar\'e invariant in the sense that the Poincar\'e generators $(H,P,J)$ are represented by Hermitian operators acting on the Hilbert space and the commutation relations exhibit no quantum anomalies. However, as we emphasized before, it is the choice of the physical coordinates (\ref{q}), (\ref{qbar}) which distinguishes this model from a free one. 
So  one may wonder whether this choice is compatible with the Poincar\'e symmetry. In particular, a natural question to ask is whether the $S$-matrix which we just derived is Poincar\'e invariant. 

To check this, let us consider a scattering process for a general total momentum $P$. Proceeding as above, let consider a general energy and momentum eigenstate  $|P,H\rangle$, with
$H>|P|>0$,
 and define the corresponding scattering wave function $\psi_{H,P}(q)$ as
\be
\langle P,q|P',H\rangle=\delta(P-P')\psi_{H,P}(q)\;.
\ee
Then, following the steps of the previous section, one arrives at the following generalization of (\ref{qeq}), 
\begin{gather}
\label{qeqP}
q\psi_{H,P}(q)={\ell_s^2\over 2}H\psi_{H,P}(q)+{2\over \pi}\int_{|P|}^\infty dH_1
\psi_{H_1,P}(q){{\cal P}\over (H_1-H)^2}\cos {\ell_s^2P(H_1-H)\over4}\;.
\end{gather}
The analysis of this equation proceeds similarly to the $P=0$ case. Namely, in the semiclassical region
\[
H-|P|\gg q^{-1}\,,\,\,(q\ell_s^2)^{-1/3}\;
\]
one finds the same wave function (\ref{psiA}) as before, where the $A_\pm$ amplitudes may now depend not only on $q$ but also on $P$. To reconstruct these amplitudes let us consider the limit
\be
\label{Plim}
 \ell_s(H-|P|)\ll 1
\ee
with $P\ell_s$ kept fixed. In this limit (\ref{qeqP}) reduces to
\be
\label{qeqP0}
q\psi_{H,P}(q)={\ell_s^2\over 2}|P|\psi_{H,P}(q)+{2\over \pi}\int_{|P|}^\infty dH_1
\psi_{H_1,P}(q){{\cal P}\over (H_1-H)^2}\;,
\ee
which is the same as (\ref{qeq0}) up to a shift of $q$ and $H$. The solution of this equation at large negative $q$ (and also accounting for (\ref{Plim})) takes the form
\be
\label{psi00}
\psi^{0}_H(q)|_{q\to-\infty}=e^{{i\over 2}q(H-|P|)+i{3\pi\over 8}}+e^{-{i\over 2}q(H-|P|)-i{3\pi\over 8}}\;.
\ee
By requiring this solution to match with the semiclassical one in the overlap region one obtains  the scattering phase shift 
\be
\label{phaseshiftP}
\delta=\ell_s^2{H^2-P^2\over 4}-{3\pi\over 4}\;,
\ee
in agreement with the Lorentz invariance of the $S$-matrix. Note that the wave function (\ref{psi00}) is the same as in the free $\ell_s=0$ theory. However, for this argument it is important that to arrive at (\ref{psi00}) we considered the limit (\ref{Plim}), resulting in (\ref{qeqP0}), rather than the naive  $\ell_s=0$ limit. This allows us to keep track of the $\ell_s^2 P^2$ term in the phase shift, as necessary for a test of  Lorentz invariance.

This result provides a non-trivial consistency check of the Lorentz invariance of our quantization. However, at the same time it raises the following puzzle. 
Namely, the momentum dependence of the wave function (\ref{psi00}) does not match (\ref{expected}).
  This issue arises already in the strict free (infinite tension) limit, $\ell_s=0$, so to understand it better let us discuss quantization of the zigzag model in this limit in more detail.

\subsection{Infinite tension limit}
In the infinite tension limit $\ell_s=0$ the expressions (\ref{q}), (\ref{p}), (\ref{qbar}) for the physical coordinates simplify to 
\begin{gather}
\label{q0}
q=-{|Q|}\\
\label{p0}
p=-{\mbox\sign(Q)\over 2}H\\
\label{qbar0}
\bar{q}=\bar{Q}\;,
\end{gather}
where, as before, at the classical level $\bar{q}$ is canonically conjugate to the total momentum $P$.
The unconventional form of the wave function (\ref{psi00}) in the asymptotic region indicates that commutators of these physical coordinates do not exhibit the canonical form with our quantization procedure.
Indeed, as follows from (\ref{qeqP0}) the $\hat{q}$ operator is defined as
 \be
\label{freeq}
\hat{q}\psi(H,P)={2\over \pi}\int_{|P|}^\infty dH_1
\psi({H_1,P}){{\cal P}\over (H_1-H)^2}\;.
\ee
It is immediate to check that this operator
does not commute with the operator 
\[
\hat{\bar{q}}=i\d_P
\]
  due to the $P$ dependence of the integration range in (\ref{freeq}). This shows that the naive expectation for the form of the position space wave function in the asymptotic region,
\[
\langle q,\bar{q}|P,H\rangle\sim e^{i {qH\over 2}+i\bar{q} P+i\delta}+h.c.\;,
\]
does not hold simply because common $q$, $\bar{q}$ eigenvectors $\langle q,\bar{q}|$  don't exist at all.\footnote{In addition to the presence of the anomaly in the $[q,\bar{q}]$, the $\bar{q}$ operator is not even symmetric with our quantization as a consequence of the $|P|<H$ constraint.} 
  This is somewhat surprising, given that the differences between quantum commutators and classical Poisson brackets are usually attributed to ordering ambiguities. At first sight these are absent for $q,\bar{q}$ as defined by (\ref{q0}), (\ref{qbar0}). However, common eigenvectors for these two operators are still absent as a consequence of the non-trivial phase space geometry. 

The states $|q,P\rangle$ considered above do exist. However, the expectation (\ref{expected}) for their asymptotic form is based on considering the matrix element of the form
\[
\langle q,P| e^{i\alpha \hat{p}}|P,H\rangle
\]
under the assumption that $\hat{p}$ acts a generator of  shifts in $q$ (and that $H=2|p|$ in the asymptotic region).  
To define the $\hat{p}$ operator based on (\ref{p0}) one needs to deal with ordering ambiguities. We did not manage to find a prescription to define $\hat{p}$ in such a way that it has a canonical commutation relation with $\hat{q}$ which is consistent with the unconventional form of the wave function (\ref{psi00}). 

In fact, applying this logic backwards, (\ref{psi00}) suggests that if we define 
\be
\label{pc1}
p_c=-{\mbox\sign(Q)\over 2}(H-|P|)\;,
\ee
then it should be possible to define the corresponding operator $\hat{p}_c$ in such a way that
\be
[\hat{q},\hat{p}_c]=i\;,\;\;[\hat{P},\hat{p}_c]=0\;.
\ee
Indeed, as we show in Appendix~\ref{app:commute} this is achieved by using the following ordering prescription  for $\hat{p_c}$,
\be
\hat{p_c}=-{1\over 2}\hat{H}^{1/2}{\sign(\hat{Q})}\hat{H}^{1/2}\;,
\ee
where the ${\sign(\hat{Q})}$ operator is defined as the Hilbert transform on a half-line
\be
\label{signQ}
\sign(\hat{Q})\psi(H,P)={2i\over \pi}\int_{|P|}^\infty dH_1
\psi({H_1,P}){{\cal P}\over H_1-H}\;.
\ee

This reasoning explains the origin of the unconventional wave function behavior exhibited in (\ref{psi00}), but may leave one wondering whether our choice of dynamical coordinates is indeed compatible with the Lorentz invariance of the system. As a self-consistency check let us present here a manifestly Lorentz-invariant quantization. It is available in the infinite tension limit and leads to the same result as above. This quantization is more conventional, in particular it operates directly in physical coordinates.

In the infinite tension limit the LR subsector of the zigzag model is described by a free Hamiltonian 
\be
\label{Hfree}
H=\left|{P\over 2}+p\right|+\left|{P\over 2}-p\right|\;.
\ee
It would still be wrong to apply the naive quantization of section \ref{sec:naive}, because the phase space still has a non-trivial geometry, which is obtained as the $\ell_s\to 0$ limit of Fig.~\ref{fig:phasespace}. Namely, the relative coordinate $q$ is restricted now to the half-line, $q<0$, and the relative momentum $p$ satisfies 
\be
\label{pfree}
|p|>{|P|\over 2}\;.
\ee
In addition,  pairs of points with opposite relative momenta $\pm p$ are identified at $q=0$. As a consequence of (\ref{pfree}) one finds that the Hamiltonian 
(\ref{Hfree}) reduces simply to
\be
H=2|p|
\ee
at all values of $P$.     Let us  quantize in the $q$, $P$ representation, so that the states are described by  wave functions $\Psi(q,P)$ with $q\in (-\infty, 0]$.
A naive quantization based on the canonical commutation relation between $q$ and $p$ fails to account for the constraint (\ref{pfree}) on the range of $p$.
Note, however, that  $p_c$ defined as
\be
\label{pc2}
p_c=p-{\sign(p)\over 2} |P|
\ee
takes values on the whole real axis.\footnote{One may worry about what happens at $p_c=0$. We ignore this issue. This is justified by the end result.}
It is straightforward to check that the definition (\ref{pc2}) agrees with the earlier one, (\ref{pc1}). Furthermore, at the level of the Poisson brackets one finds 
\be
\{q,p_c\}=1-{1\over 2}\delta(p)|P|=1\;,
\ee
where at the last step we made use of (\ref{pfree}).
On the other hand, we now have
\[
\{\bar{q},p_c\}=-{1\over 2}\sign(p)\sign(P)\;.
\]
This can be fixed by introducing
\be
\label{qbarc}
\bar{q}_c=\bar{q}+q \sign(p)\sign(P)\;.
\ee
Then one finds  that $(q,p_c,\bar{q}_c, P)$ form  a set of canonical variables on the classical phase space at hand,
\begin{gather}
\{\bar{q}_c,p_c\}=\{q,P\}=\{p_c,P\}=0\\
\{\bar q_c, P\}=\{q,p_c\}=1\\
\{\bar{q}_c,q\}=q\delta(p)\sign(P)=0\;,\label{bqq}
\end{gather}
where in evaluating the last Poisson bracket (\ref{bqq}) we again made use of (\ref{pfree}). Given that in terms of these variables the 
only constraint we have is $q\leq 0$, it is natural use these for a canonical quantization, {\it i.e.}, to define
\[
\hat{\bar{q}}_c=i\d_P\;,\;\; \hat{p}_c=-i\d_q\;.
\]
The Hamiltonian (\ref{Hfree}) takes the following form in these variables,
\be
H=2|p_c|+|P|\;.
\ee
Following the same procedure to define $|\hat{p_c}|$ as before, we conclude  that wave functions $\Psi(q,P)$ satisfy the Schr\"odinger  equation with the Hamiltonian $\hat{H}$
given by
\be
\label{hatH}
\hat{H}\Psi(q,P)=|P|\Psi(q,P)-{2\over \pi}\int_{-\infty}^0dq_1\Psi(q_1, P){{\cal P}\over (q_1-q)^2}\;.
\ee
The functional form of the corresponding energy eigenstates $\Psi_H(q,P)$ is
\be
\Psi_H(q,P)=f(q(H-|P|))\;.
\ee
 On the other hand, as follows from (\ref{freeq}), the $\bar{q}$ eigenfunctions $\psi_q(H,P)$ in the quantization we had before take exactly the same form,
 \be
 \psi_q(H,P)=f(q(H-|P|))\;
 \ee
with the same function $f$ (whose explicit form can be found in Appendix~\ref{app:Mellin}). Hence, in the infinite tension limit the quantization relying on dynamical coordinates is equivalent
to the one based on the conventional Schr\"odinger equation.

Let us check now that the Schr\"odinger quantization is also manifestly Lorentz invariant. Classically, the boost generator (\ref{boost}) takes the following form in the $(q,p_c,\bar{q}_c, P)$ variables,
\be
\label{Jfree}
J=\bar{q}_c(2|p_c|+|P|)-2\sign(P) qp_c\;.
\ee
Just like before, a normal ordering ambiguity cancels out between two terms in (\ref{Jfree}), so at the quantum level we may define
the boost operator as
\be
\label{Jhatfree}
\hat{J}\Psi(q,P)=i\d_P\l|P|\Psi(q,P)- {2\over \pi}\int_{-\infty}^0dq_1\Psi(q_1, P){{\cal P}\over (q_1-q)^2}\r+i \sign(P)q\d_q\Psi(q,P)\;.
\ee
It is straightforward to check now that $\hat{J}$, $\hat{H}$ and $\hat{P}$ form the Poincar\'e algebra $ISO(1,1)$.

\section{Multi-particle case}
\label{sec:mult}

Generalizing the two-particle discussion, the zigzag solution for $N$ particles can be recast as the dynamics of $N$ free particles with restricted momenta. Each sector with distinct topological charge $T = N_L-N_R$ leads to $N_L$ free left-movers and $N_R$ free right-movers. Amongst each set (being left or right-movers) there is an absolute ordering analogous to the LL (RR) sectors of the two-particle case. Below we explain in detail how this picture arises. 

In the construction of \cite{Donahue:2019fgn}, almost all integrals of motion are functions that have support along diagonals in the string ``bit space'' that intersect the physical region defined by the topological charge. Asymptotically, each of these integrals reduces to either a momentum $p_i$ or a coordinate difference $q_i-q_{i+1}$ between only left- or only right-movers. To construct free particle variables we call these integrals $P_i$ and $Q_i - Q_{i+1}$ respectively. Since the particles in the zigzag model have definite asymptotic ordering, $Q_i - Q_{i+1}$ integrals are all of a definite sign in the topological sector in which they are defined. The same is true for the $P_i$ integrals whose sign determines whether a particle is a left- or right-mover asymptotically. 
Therefore we arrive at $N_L$ free left-movers and $N_R$ free right-movers that have definite orderings amongst themselves. In total we have $N_R+N_L$ integrals $P_i$, $N_L-1$ integrals $Q^L_{i}- Q^L_{i+1}$, and $N_R-1$ integrals  $Q^R_{i}- Q^R_{i+1}$. To get canonical pairs $(P_i,Q_i)$, we need to supplement this set with variables $Q_L-Q_R$ and $Q_L+Q_R$. 

The ``clock" variable, $\tilde{H}$, in this picture is schematically a coordinate like $Q_L-Q_R$ that increases linearly in time as one would expect for free particles. As defined, however, it is mixed with some other integrals which we'll need to subtract to obtain canonical Poisson brackets. This was also the case in the two-particle solution discussed in section \ref{sec:interactingLR}, as is clear by subtracting the two definitions in (\ref{eq:freepos})
\be
\label{eq:diff}
Q_1-Q_2 = \tilde{H} - \frac{1}{2}(P_2 + P_1) \,\,\, .
\ee
The recipe for $Q_R-Q_L$ in the $N$ particle sector will then just be a generalization of (\ref{eq:diff}). If we choose to label the free particles such $Q^R_1< \ldots < Q^R_{N_R}$ and $Q^L_1< \ldots < Q^L_{N_L}$, then 
analogously to (\ref{eq:diff}) we define
%
\be
\label{eq:Ndiff}
Q^R_{N_R} - Q^L_{N_L} \equiv \tilde{H} - \frac{1}{2} P \,\,\, .
\ee

Then, as in the two particle case, we can use the fact that $Q^L_i$ and $Q^R_j$ transform with opposite signs under the action of boosts to define the individual positions
\begin{gather}
    Q^L_{N_L} \equiv -\frac{1}{2}( Q^R_{N_R}- Q^L_{N_L} + \{J,Q^R_{N_R}- Q^L_{N_L}\} ) \\
    Q^R_{N_R} \equiv \frac{1}{2}(Q^R_{N_R}- Q^L_{N_L} - \{J,Q^R_{N_R}- Q^L_{N_L}\} ) \,\,\,  .
\end{gather}
In terms of the last integral of motion $\tilde{P} = \{\tilde{H},J\}$ defined in \cite{Donahue:2019fgn}, we have 
\be
Q^R_{N_R} + Q^L_{N_L} = \tilde{P} + {1\over2}H.
\ee
 All together we have $2(N_R+N_L)-1$ integrals of motion, as this system is classically maximally superintegrable. Since the Poisson brackets of integrals again give integrals, we may calculate all Poisson brackets in the asymptotic region and extrapolate to the whole topological sector of phase space. In fact, this trick also works for Poisson brackets with $Q^L_{N_L}$ and $Q^R_{N_R}$ since their Poisson brackets with $H$ are constant. Given the explicit construction of \cite{Donahue:2019fgn}, we know the asymptotic form of these integrals and we find the canonical brackets one would expect from our suggestive naming scheme
\begin{align}
    \{Q^R_i, P^R_j\} & =\delta_{ij} \\
    \{Q^L_i, P^L_j\}&  =\delta_{ij} \,\,\, 
\end{align}
with all other brackets being zero. As in a two particle case  we find that the expressions for the Lorentz algebra can be rewritten into the form expected for free particles
\begin{gather}
\label{eq:multi}
    H = \sum_i P^R_i - \sum_i P^L_i \,\, , \,\, P = \sum_i P^R_i + \sum_i P^L_i \\
    J =  \sum_i Q^R_i P^R_i - \sum_i Q^L_i P^L_i \,\,\, .
\end{gather}

We are led to conclude that the zigzag solution for $N$ particles can be recast as the dynamics of $N$ free particles with restricted momenta and positions. Each sector with distinct topological charge $T = N_L-N_R$ leads to $N_L$ strictly left movers and $N_R$ strictly right movers. Amongst each set (being left or right-movers) there is also an ordering $Q_i - Q_{i+1} < 0$. To preserve this structure its natural to quantize the set of coordinates $(Q_i^R,P_i^R)$ and $(Q_i^L,P_i^L)$, with phase space $ (\mathbb{R} \times \mathbb{R}^+)^{N_R} \times (\mathbb{R} \times \mathbb{R}^-)^{N_L}$. As in the two particle $LL/RR$ sectors, we enforce the constraint $Q^{R(L)}_i - Q^{R(L)}_{i+1} < 0$ by imposing boson/fermion statistics amongst right- (left-) movers. Just as before, this entails a non-trivial modification of the zigzag model but is natural from the view of the parent adjoint $QCD_2$.

The $2N-1$ integrals of motion select a unique trajectory in phase space and the final ``clock" $Q^R_{N_R}-Q^L_{N_L}$ chooses a point on this trajectory. Clearly this map is one-to-one and as before one may construct the inverse maps $q_i (Q_i,P_i)$ and $p_i (Q_i,P_i)$. With these definitions in hand and following the procedures of the previous sections, in  principle one should be able to extract quantities such as the S-matrix. Of course this quickly becomes cumbersome to perform in detail, so we will not pursue such an investigation here. On the other hand, it is straightforward to follow the semi-classical argument presented at the end of section \ref{subsec:restframe} to derive the momentum dependence of the time delay, which again reproduces the $T\bar{T}$ phase shift
(\ref{shock}).


\section{Comments on (Folded) Closed Strings}
\label{sec:thooft}
\begin{figure}
    \centering
    \textbf{}\par\medskip
    \includegraphics[scale=0.5]{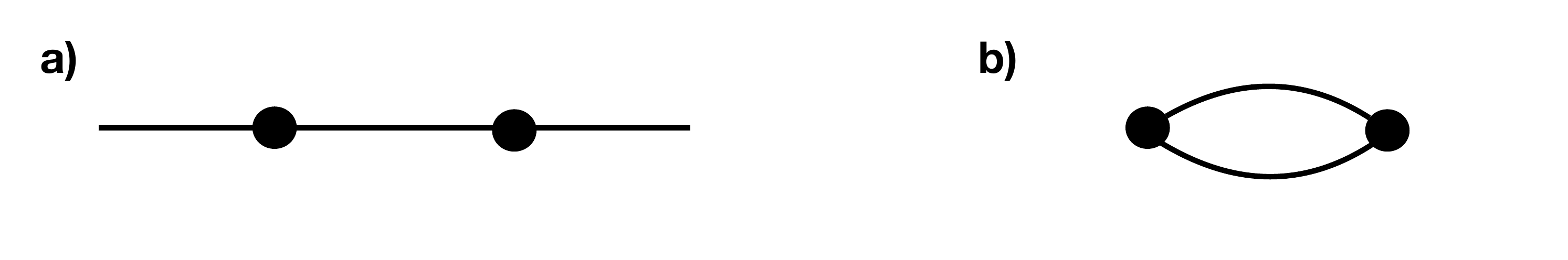}
    \caption{A long string configuration with two partons, corresponding to the zigzag model a),  and a short closed ``folded'' string with two partons b).}
    \label{fig:zigzags}
\end{figure}

The zigzag model describes high energy dynamics of a long string in adjoint QCD$_2$. It is natural to also consider its closed string analogue, see Fig.~\ref{fig:zigzags}. This is the ``folded string" model introduced back in \cite{Bardeen:1975gx} (see \cite{Xi} for a recent overview).
Restricting to massless quarks and to a two-particle subsector the latter is given by the following Hamiltonian
\be
\label{2pclosed}
H = |p_1|+|p_2|+2|q_1-q_2| \,\,\, .
\ee
Both models can be obtained from the action describing the two-dimensional  Nambu--Goto string interacting with massless point particles,
\[
S=S_{NG}+S_{pp1}+S_{pp2}\;,
\]
where
\[
S_{NG}=-\ell_s^{-2}\int d^2\sigma\sqrt{-\det\d_\alpha X^a\d_\beta X^b}=-{1\over 2}\ell_s^{-2}\int d^2\sigma\epsilon^{\alpha\beta}\epsilon_{ab}\d_\alpha X^a\d_\beta X^b
\]
and
\[
S_{pp}=\int d\tau e (\d_{\tau} X^a)^2\;.
\]
The difference between two sectors is entirely due to different choices of how the strings are attached to the particles, as illustrated in Fig.~\ref{fig:zigzags}.
It is natural to ask whether the analysis of the long string sector presented above teaches us anything about the closed string sector.
Probably the main lesson we learned so far is that a consistent quantization of these models requires a careful accounting for the phase space geometry. Indeed, it was observed back in \cite{Artru:1983gm,Lenz:1995tj} that a straightforward quantization of the Hamiltonian (\ref{2pclosed}) based on  the canonical commutation relations (\ref{canonical}) is inconsistent with the Poincar\'e invariance of the model and does not lead to a boost-invariant mass spectrum of closed strings. This is very similar to the situation we encountered in section~\ref{sec:naive}. 

\begin{figure}
    \centering
    \textbf{}\par\medskip
    \includegraphics[scale=0.5]{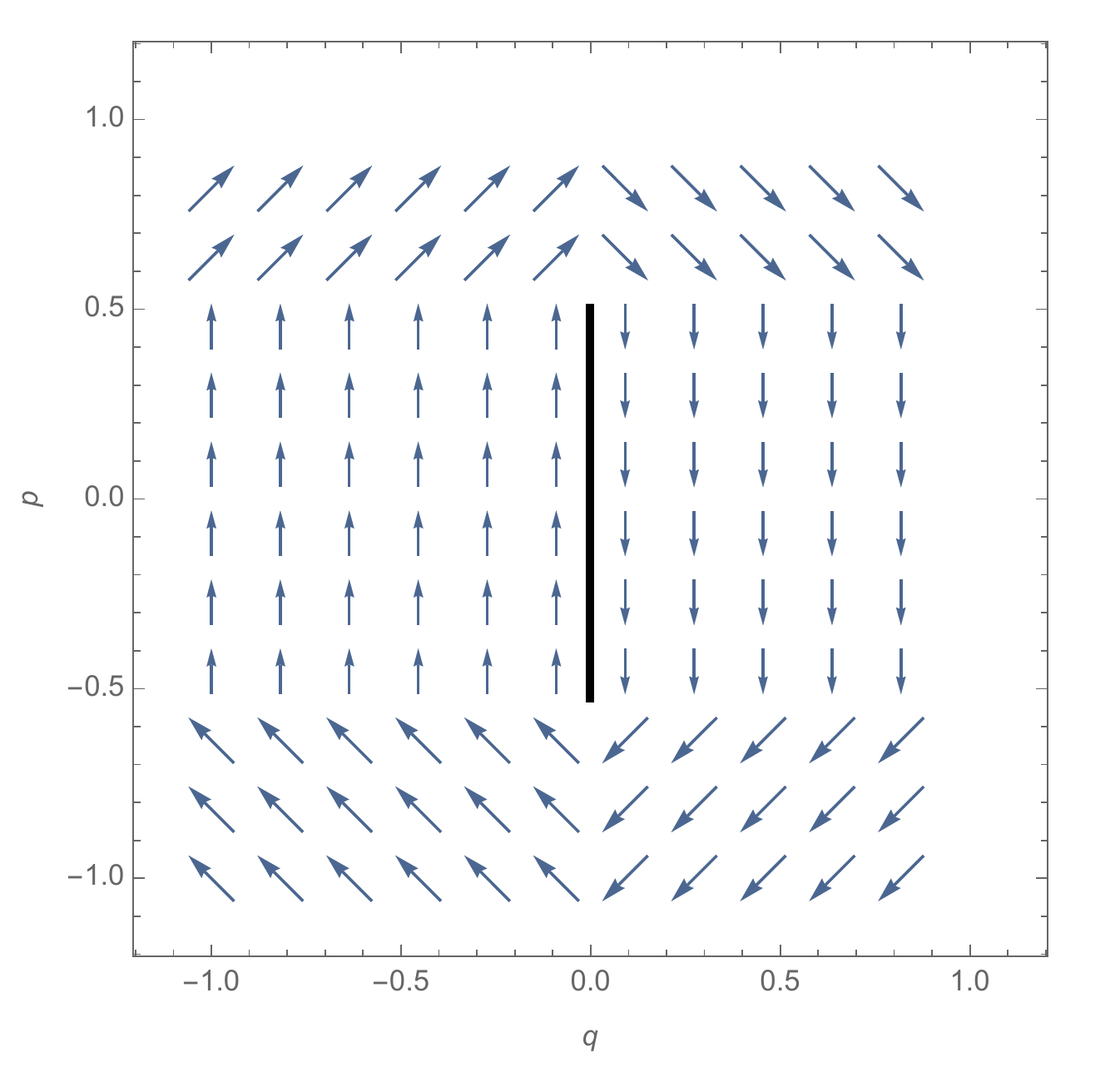}
    \caption{The folded  string Hamiltonian flow in the $(p,q)$ plane for the total momentum $P=1$.  A solid line shows the discontinuity of the flow, which should be excluded from the phase space.}
    \label{fig:thooft}
\end{figure}

To see that this problem has a similar origin it is instructive to inspect the phase portrait of  (\ref{2pclosed}). In Fig.~\ref{fig:thooft} we presented an analogue of 
Fig.~\ref{fig:phasespace}. We observe that short closed strings describe a single topological sector. However, similarly to the zigzag model, the Hamiltonian flow exhibits a bad discontinuity along the interval $q=0$, $|p|<|P|/2$ in the $(q,p)$ plane. This strongly suggests that a consistent quantization of the closed string sector should be performed by excluding this interval from the phase space.
This leads us to a phase space which has a topology of $\mathbb{R}^2\times \mathbb{R}_+\times {\cal S}^1$ rather than simply $\mathbb{R}^4$. Of course, the non-trivial part of the problem is the quantization of $(q,p)$ variables, which correspond to the $ \mathbb{R}_+\times {\cal S}^1$
part of the phase space, which is topologically equivalent to a plane with an excluded point. 

Following the logic of section~\ref{sec:Zigzag} it is natural to attempt to quantize closed string sector using the action-angle variables. For a conventional 
one-dimensional system with $ \mathbb{R}^2$ phase space this quantization is problematic (see, e.g., \cite{Susskind:1964zz}), because the angle variable is not globally well-defined even at the classical level. This obstruction is not present for the  $ \mathbb{R}_+\times {\cal S}^1$ phase space topology. Quantization in the action-angle variables implies that the Bohr--Sommerfeld conditions
\[
\oint pdq =2\pi n
\]
determine energy levels exactly as a consequence of the periodicity of the angle variable  (up to a possible constant shift related to a magnetic flux piercing the plane through the origin). This results in the exactly linear Regge trajectory for closed strings,
\be
H^2-P^2=\pi n+const\;.
\ee

We feel that there are several reasons to be cautious about self-consistency of this quantization. First, another plausible expectation for the closed string spectrum follows from the 't Hooft equation \cite{tHooft:1974pnl} for large $N$ QCD with fundamental quarks. It provides yet another quantization for (\ref{2pclosed}). Interestingly, the resulting spectrum is very close to the linear Regge trajectory, and deviations from the exact linearity are at the percent level even for the lowest lying states \cite{Fateev:2009jf}. Still, these deviations are non-zero. 

More generally,  the only specific proposal for quantization of $ \mathbb{R}_+\times {\cal S}^1$  we are aware of is the one put forward in\cite{Bojowald:1999pf}.
It is based on the $SL(2,R)$ action on $ \mathbb{R}_+\times {\cal S}^1$ and leads indeed to the exactly linear Regge trajectory. However, it appears that 
the resulting $SL(2,R)$ representations actually correspond to the geometric quantization of the space-like hyperbolic  coadjoint $SL(2,R)$ orbits \cite{Witten:1987ty}, which have a topology of the disc.
The coadjoint orbit of $SL(2,R)$ with the $ \mathbb{R}_+\times {\cal S}^1$ topology is the (null) cone, and to the best of our knowledge its quantization is unknown.

These considerations suggest that the straightforward action angle quantization of closed strings is missing a subtle quantum effect. If so, this situation would be similar to what happens for strings in $D=3$ space-time dimensions. 
In that case, an integrable quantization is consistent in the long string sector and is given by the $T\bar{T}$-deformation \cite{Dubovsky:2015zey}. However,
its short string analogue, which is a light cone quantization, suffers from a global Poincar\'e anomaly leading to the presence of irrational anyons in the spectrum
\cite{Mezincescu:2010yp}. 
Alternatively, it is also possible that the system (\ref{2pclosed}) and its mulitparticle generalizations admit several inequivalent consistent quantizations (c.f. \cite{Kalashnikova:1997cs}), and one of them corresponds to the linear Regge trajectory. We leave the study of this interesting question for the future.

\section{Discussion}
\label{sec:last}
To summarize, in this paper we described how to quantize the zigzag model consistently with  Poincar\'e symmetry and integrability. It appears that the principal lesson to draw from our results is that a consistent quantization of this  model requires a careful accounitng for the non-trivial geometry of the phase space. This lead us to the quantization with the expected properties---Poincar\'e invariance and a quantum phase shift which exactly reproduces the classical time delay. We feel, however, that this study is only a first step towards the proper understanding of the quantum zigzag model. Indeed, the integrable structure of the classical zigzag model allows for a very elegant and suggestive formulation in terms of the discrete geometry of the ``classical bit space" \cite{Donahue:2019fgn}. This gives rise to a hope that a comparably elegant description of the quantum zigzag model should be possible. We don't  think this was achieved in the current work. Apart from purely aesthetic reasons, there is also a practical motivation to look for an improved description of the quantum zigzag model. Namely, the original motivation for our study was to use this model as a basis for high energy expansion on the worldsheet of confining strings in  two-dimensional adjoint QCD. However, this goal looks quite hard to achieve using the formalism presented here. 

A very interesting property of the zigzag model  is that it leads to the shock wave phase shift (\ref{shock}), which also describes massless $T\bar{T}$ deformed theories. However, it looks likely that the physics of the zigzag model is somewhat different. One indication  comes from the fact that the full $S$-matrix which we obtained 
in the zigzag model (\ref{phaseshiftP}) contains an additional constant  $-3\pi/4$ phase shift. This phase shift is well familiar from the semiclassical analysis of the 't Hooft equation  \cite{Brower:1978wm,Fateev:2009jf} and does not have an analogue in the $T\bar{T}$ case. Furthermore, even though the two models lead to  identical time delays, the underlying physical mechanism is quite different. The  $T\bar{T}$  time delay may be understood  as coming from the fact that the proper length of the perturbed string worldsheet stretches proportionaly to the excitation energy \cite{Dubovsky:2012wk}. As a result the $T\bar{T}$ scattering always corresponds to the total transmission with time delays caused by the above stretching.  
On the other hand, in the zigzag case the time delay is caused by a zigzag string configuration resulting in total reflection. 
This is incompatible with integrability for particles of different masses, unlike for the $T\bar{T}$ deformation which exists for arbitrary masses of colliding particles. 
It will be interesting to understand better the relation between  the two models (see \cite{Cardy:2020olv} for similar ideas).
We hope to address these and other related questions in the future.

{\it Acknowledgements.}
We thank Alexander Artemiev, David Gross, Shota Komatsu, Ivan Kostov and Fedor Popov for discussions. This work has been completed during the KITP program on Confinement, Flux Tubes and Large $N$. We thank all participants of the program for creating a very stimulating environment.
 This research was supported in part by the National Science Foundation under grants No. NSF PHY-1748958 and PHY-1915219, and by the BSF grant 2018068.
 
\appendix
\section{ Numerics for the naive phase shift}
\label{app:numerics}
The naive quantization results in the Schr\"odinger equation (\ref{eq:SE}) of the hypersingular form. Solving it numerically  directly in the position space looks somewhat problematic due to a singularity present in the integral kernel. However, a very efficient way to deal with this kernel is to make use of the Fourier transform. Namely we start with
a spatial grid of $N_p$ points in a finite spatial box $q\in (-L,L)$. We evaluate the integral term of (\ref{eq:SE}) by first performing the (Fast) Fourier transform, then by multiplying the result by $|p|$ and finally by performing the inverse Fourier transform. The potential term is evaluated directly in the position space. Then one finds the eigensystem of the resulting discretized Hamiltonian. The phase shift is found by evaluating the numerical derivative of the resulting eigenfunctions deep in the free region $q<0$,
\[
\delta(p)=2\l i\tan^{-1}\l{1\over p}{\d_q\psi\over \psi}\r-pq\r+const\;.
\]
 The phase shift obtained by implementing this procedure in Mathematica with $L=50$ and $N_p=4000$
 is presented in Fig.~\ref{fig:phase}. 
 
 As a cross-check we also determined the phase shift by directly time evolving a narrow initial wave packet using the time-dependent Schr\"odinger 
 equation (\ref{eq:SE}) (again evaluating the kernel using the Fourier transform). This method is less accurate and harder to implement (although, it works better for the equation  (\ref{eq:SE}) than for a conventional Schr\"odinger equation, because wave packets keep their shape constant in the free region for  (\ref{eq:SE})). Nevertheless, we obtained the agreement between these two methods, which is good enough to be confident that the deviation of the phase shift shown in Fig.~\ref{fig:phase} from the shock wave one is real and trustworthy.
 
\section{Solving the scattering equation in the infinite tension limit}
\label{app:Mellin}
 
In the infinite tension limit, the eigenvalue problem for the $\hat{q}$ operator reduces to solving the following equation
\be
\label{freeqeq}
q\psi(H,P)={2\over \pi}\int_{|P|}^\infty dH_1
\psi({H_1,P}){{\cal P}\over (H_1-H)^2} \,\,\, .
\ee
This equation has appeared before in the scaling limit of the 't Hooft equation with zero renormalized quark mass \cite{Brower:1978wm,Fateev:2009jf}, and more recently in studies of fractional laplacians in bounded domains \cite{kuznetsov2018spectral,kulczycki2010spectral}. The solution to this equation is most transparently presented in Mellin space, via the transform
\be
 \phi(\lambda)=[\mathcal{M}\psi](\lambda)=\int_0^\infty dK K^{\lambda-1} \psi(K) \,\,\, ,
\ee
where $K\equiv H-|P|$ and we suppressed the dependence on $P$. Our eigenvalue problem reduces to the difference equation
\be
\label{eq:diffeq}
    q\phi_q (\lambda)= -2 (\lambda-1) \cot(\pi \lambda) \phi_q (\lambda-1)    \,\,\, ,
\ee
with $\hat{q} \phi_q (\lambda) = q\phi_q (\lambda)$.  Solutions to this equation are straightforward to generate with the {double-sine function} $S_2(\lambda;b)$ in hand. In the notation of \cite{kuznetsov2018spectral}, it is defined by the relations 
\be
S_2(\lambda+1;b) = \frac{S_2(\lambda;b)}{2 \sin{\pi \lambda /b}}, \hspace{1cm} S_2(\lambda+b;b)= \frac{S_2(\lambda;b)}{2 \sin{\pi \lambda}} \,\,\, ,
\ee
see \cite{Volkov_2005} for further comments, we set $b=1$ and suppress it in what follows. Solutions to eq. (\ref{eq:diffeq}) are of the form
\be
    \phi_q (\lambda) = \Big(\frac{-2}{q}\Big)^{\lambda} P(\lambda)  \frac{\Gamma(\lambda)S_2(\lambda)}{S_2(\lambda+1/2)} \,\,\,,
\ee
where $P(\lambda)=P(\lambda+1)$ is an arbitrary periodic function. Note that $\phi(\lambda)$ is only exponentially bounded for $q<0$, therefore the spectrum of $\hat{q}$ is $\mathbb{R}^-$. Requiring that $\phi(\lambda)$ is bounded as $\mbox{\rm Im}(\lambda) \to \infty$ and analytic in $0<\mbox{\rm Re}(\lambda)<2$ restricts $P(\lambda)=1$. The asymptotic behaviour of the double-sine functions is given by  \cite{kuznetsov2018spectral}
\be
S_2(\lambda) \backsim 
\begin{cases}
e^{\frac{i \pi}{2}(\lambda^2-2\lambda)} \hspace{1cm} \,\,\, \mbox{\rm Im}(\lambda) \to \infty \\
e^{-\frac{i \pi}{2}(\lambda^2-2\lambda)} \hspace{1cm} \mbox{\rm Im}(\lambda) \to -\infty 
\end{cases}\,\,\, .
\ee
Along with the well-known asymptotics of the Gamma function, a saddle-point analysis yields
\be
\psi_q(K) \equiv \frac{1}{2 \pi i}  \int_{\mathcal{C}} d\lambda K^{-\lambda} \phi_q(\lambda) \xrightarrow[qK \to \infty]{}  e^{i q K/2} e^{3\pi i/8}+e^{-i q K/2} e^{-3\pi i/8} \,\,\,,
\ee
as quoted in the main text. An explicit expression for the remainder $r(qK) \equiv \sin(qK/2+\pi/8)-\psi_q(K) $ appears in \cite{kulczycki2010spectral}. It is straightforward to formulate the interacting ($l_s \neq 0$) COM-frame eigenvalue problem in Mellin space as well
\be
     q\phi_q (\lambda)= -2 (\lambda-1) \cot(\pi \lambda) \phi_q (\lambda-1) + \frac{1}{2\alpha} \phi_q(\lambda+1)\,\,\, .
\ee
Despite some effort, we have not been able to solve this equation.

 \section{  $p_c$ commutation relations}
 \label{app:commute}

Now we would like to show that $[\hat{q},\hat{p_c}]=i$ if we start with $[\hat{Q},\hat{K}]=i$. For convenience we introduce the following two operators,
\be
\tilde{\epsilon}(\hat{Q}) \psi(K) \equiv \frac{-i}{\pi \sqrt{K}} \dashint_{0}^\infty dK'  \sqrt{K'} \frac{\psi(K')}{(K'-K)} 
\ee
and
\be
\bar{\epsilon}(\hat{Q}) \psi(K) \equiv \frac{-i\sqrt{K}}{\pi } \dashint_{0}^\infty dK'   \frac{1}{\sqrt{K'}}\frac{\psi(K')}{(K'-K)} 
\ee
which are conjugate to one another. Furthermore, these are both inverses to the well known Hilbert transform
\be
\epsilon(\hat{Q}) \psi(K) \equiv \frac{-i}{\pi } \dashint_{0}^\infty dK'   \frac{\psi(K')}{(K'-K)} 
\ee
which is self-adjoint and bounded. Unlike on the real line, where its eigenvalues are the points $\pm 1$, on the half-line the spectrum of the Hilbert transform is the continuum $\sigma\in[-1,1]$ \cite{koppelman1959spectral}. Indeed for $-1<\mu<0$ we have
\be
\epsilon(\hat{Q}) K^{\mu} = i \cot(\pi \mu) K^{\mu}
\ee
and its generalized eigenfunctions are $K^{-1/2+ir}$ with $\sigma = \mbox{tanh}(\pi r)$. It's inverse is clearly $\sigma^{-1}$, but this demands an $i \epsilon$ procedure to avoid the pole at $r=0$. The two signs for this $i \epsilon$ give rise to the two operators $\bar{\epsilon}$ and $\tilde{\epsilon}$, where we suppress $\hat{Q}$ in what follows. In analogy with the propagators of field theory, they give a basis for generic Green's functions. From this discussion it is clear that
\begin{align}
    \epsilon \tilde{\epsilon} =\tilde{\epsilon} \epsilon  =  1 = \epsilon \bar{\epsilon} =\bar{\epsilon} \epsilon
\end{align}
With this compact notation, we have $\hat{p_c}=-K\tilde{\epsilon} =- \bar{\epsilon}K$ and $\hat{q}=-i\d_K \epsilon$, and the commutator is
\begin{align}
\label{132}
    [\hat{q},\hat{p_c}]& =i[\d_K \epsilon, K \tilde{\epsilon}]=i\d_K \epsilon  K \tilde{\epsilon}-iK \tilde{\epsilon} \d_K\epsilon = i\d_K \epsilon  \bar{\epsilon} K -iK \tilde{\epsilon}\epsilon \d_K\\
    &     = i\d_K  K -iK  \d_K =i\;.
\end{align}
When integrating by part in the second term in (\ref{132}) we generate a boundary term
\be
i\d_K \epsilon \psi =  \frac{1}{\pi } \dashint_{0}^\infty dK'   \d_K\frac{1}{(K'-K)} \psi(K') =\frac{1}{\pi } \dashint_{0}^\infty dK'   \frac{\d_{K'} \psi(K')}{(K'-K)}  + \frac{\psi(0)}{\pi K}
\ee
which is subsequently annihilated by $\tilde{\epsilon}$ as from above we have $\epsilon K^{-1/2} = 0 = \tilde{\epsilon} K^{-1}$. 

\bibliographystyle{utphys}
\bibliography{dlrrefs}
\end{document}